\pdfoutput=1
\documentclass[useAMS,usenatbib]{mn2e}
\usepackage{graphicx}
\usepackage{amssymb}
\usepackage{lscape}
\usepackage{rotating}
\usepackage{subfigure}
\usepackage{amsmath}

\def\deg{\ifmmode^\circ\else$^\circ$\fi}

\def\Q{\ifmmode\mathcal{Q}\else$\mathcal{Q}$\fi}
\def\Mach{\ifmmode\mathcal{M}\else$\mathcal{M}$\fi}

\title[New candidate hub-filament system in G25.4NW]
{Unraveling the inner substructure of new candidate hub-filament system in the HII region G25.4NW}

\author[L.~K. Dewangan]
{L.~K. Dewangan$^{1}$\thanks{lokeshd@prl.res.in}\\
$^{1}$Physical Research Laboratory, Navrangpura, Ahmedabad - 380 009, India.\\}

\begin{document}

\date{ }

\pagerange{\pageref{firstpage}--\pageref{lastpage}} \pubyear{2020}

\maketitle

\label{firstpage}

\begin{abstract}
We present multi-scale and multi-wavelength data of the Galactic H\,{\sc ii} region G25.4-0.14 (hereafter G25.4NW, distance $\sim$5.7 kpc). The SHARC-II 350 $\mu$m continuum map displays a hub-filament configuration containing five parsec scale filaments and a central compact hub. Through the 5~GHz radio continuum map, 
four ionized clumps (i.e., Ia--Id) are identified toward the central hub, and are powered by massive OB-stars. 
The {\it Herschel} temperature map depicts the warm dust emission (i.e., T$_{d}$ $\sim$23--39~K) toward the hub. 
High resolution Atacama Large Millimeter/submillimeter Array (ALMA) 1.3 mm continuum map (resolution $\sim$0\rlap.{$''$}82 $\times$ 0\rlap.{$''$}58) reveals three cores (c1--c3; mass $\sim$80--130 M$_{\odot}$) toward the ionized clumps Ia, and another one (c4; mass $\sim$70 M$_{\odot}$) toward the ionized clump Ib. A compact near-infrared (NIR) emission feature (extent $\sim$0.2 pc) is investigated toward the ionized 
clump Ia excited by an O8V-type star, and contains at least three embedded K-band stars.
In the direction of the ionized clump Ia, the ALMA map also shows an elongated feature (extent $\sim$0.2 pc) hosting the cores c1--c3. All these findings together illustrate the existence of a small cluster of massive stars in the central hub. 
Considering the detection of the hub-filament morphology and  the spatial locations of the mm cores, a global non-isotropic collapse (GNIC) scenario appears to be applicable in G25.4NW, which includes the basic ingredients of the global hierarchical collapse and clump-fed accretion models. Overall, the GNIC scenario explains the birth of massive stars in G25.4NW. 
\end{abstract}
\section{Introduction}
\label{sec:intro}
Massive OB-type stars (M $\gtrsim$ 8 M$_{\odot}$), through ultraviolet (UV) radiation, stellar winds, radiation pressure and supernova explosions, have shaped the formation and evolution of galaxies. However, the problem of massive star formation (MSF) is still unsettled \citep[e.g.,][]{zinnecker07,tan14,Motte+2018,hirota18,rosen20}.  
Concerning MSF, the process of mass accumulation through inflow from larger scales has recently received significant attention \citep{rosen20}. In particular, a hub-filament system has been thought as the important observational site, where several parsec-scale filaments intersect the central denser regions  
\citep[e.g.,][and references therein]{myers09,schneider12,Motte+2018,morales19,kumar20,dewangan20x}. On the basis of the hub-filament system, an evolutionary scheme for MSF \citep[i.e., a global non-isotropic collapse scenario (GNIC); see][]{Motte+2018,Tige+2017} is proposed. 
In this scenario, infrared bright massive protostars grow from low mass stellar embryos, which gain the material through gravitationally-driven inflows from the parental massive dense cores/clumps \citep[MDCs, in a 0.1~pc scale; see Figure~8 in][]{Tige+2017}. These MDCs are firstly developed by a large amount of inflowing gas through filaments. In this relation, one needs to identify the embedded filaments along with the central hub and to explore the inner environments of the central hub hosting massive OB-type stars. 
In this context, the target of this paper is a prominent Ultra Compact (UC) H\,{\sc ii} region GPSR5 25.398-0.141 or G25.4-0.14 (hereafter G25.4NW).

Situated at a near-side distance of 5.7 kpc, G25.4NW lies in the inner Galactic molecular ring region, and is excited by 
at least an O6-type star \citep{zhu11,ai13}.  
The molecular gas toward G25.4NW was studied in a velocity range of [90, 101] km s$^{-1}$ \citep{ai13}. 
G25.4NW is associated with a dust continuum source (or clump) JCMT 18354-0649N (hereafter J18354N) \citep{lester85}, which is 
located about 1$'$ north of other dust continuum source JCMT 18354-0649S (hereafter J18354S) \citep[see Figure~1 in][]{carolan09}. 
It was reported that the clump J18354S has a more compact appearance than the clump J18354N in the SCUBA dust continuum maps at 450 and 850 $\mu$m \citep{wu05}. 
\citet{zhu11} presented sub-arcsecond resolution H-, K-, and L$'$-band images toward both the continuum sources, which were observed using the United Kingdom Infrared Telescope (UKIRT) facility. 
The near-infrared (NIR) nebulosity was seen toward the central part of J18354N, where at least three stars were observed \citep[see Figure~3 in][]{zhu11}. However, there are no dust continuum maps with sub-arcsecond 
resolution reported toward G25.4NW. But, such observations were presented toward the clump J18354S (see \citet{liu11} and also Figure D.6 in \citet{zhang19}). 
A green fuzzy object or an extended green object (EGO) has been identified toward J18354S using the {\it Spitzer} infrared images \citep[see Figure~1 in][]{carolan09}, where inflow and outflow motions were investigated \citep{zhu11,zhang19,zhang20}. 
In the direction of the EGO buried in the source J18354S, \citet{zhu11} identified an embedded source (i.e., IRS1a) in the UKIRT K-band image that was characterized 
as a massive protostar \citep[M $\sim$6--12 M$_{\odot}$; see Figure~3 in][]{zhu11}. It was also suggested that the massive protostar is forming through rapid accretion \citep{wu05,liu11,zhu11,zhang19,zhang20}. 
However, the formation process of the massive O-type star in G25.4NW is yet to be explored. 

In the literature, there is no study available using the high resolution sub-millimeter (sub-mm) or millimeter (mm) continuum data 
toward G25.4NW, which can be used to examine the presence of filaments and cores. 
In this paper, we aim to understand the physical process related to the birth of the O-type star in G25.4NW, which also 
includes understanding of the mechanism of mass accumulation in MSF. 
In this context, we employ multi-scale 
and multi-wavelength data sets of the continuum source JCMT 18354N or G25.4NW, which include the high 
resolution 5 GHz radio continuum map and the Atacama Large Millimeter/submillimeter Array (ALMA) continuum maps. 
Furthermore, we revisit several published multi-wavelength data of G25.4NW for exploring its physical environment. 
In Section~\ref{sec:obser}, we list various observational data sets utilized in this paper. 
Section~\ref{sec:data} presents observational outcomes at different scales and different wavelengths toward G25.4NW. 
It includes the identification of filaments, ionized clumps, and dust continuum sources toward G25.4NW. 
In Section~\ref{sec:disc}, we discuss the implications of our observed outcomes.  
Finally, Section~\ref{sec:conc} summarizes the major findings.
\section{Data sets}
\label{sec:obser}
The paper utilizes several observational data sets from available surveys, which are listed in Table~\ref{tab1}. 
The data sets were downloaded toward an area of $\sim$6\rlap.{$'$}6 $\times$ 4\rlap.{$'$}2 containing G25.4NW, 
which is centered at {\it l} = 25$\degr$.39; {\it b} = $-$0$\degr$.135 (Figure~\ref{fig1}). 
In this paper, our study exclusively focuses on the source J18354N or G25.4NW. 
Hence, we do not discuss the observational results of the well studied source J18354S. 

Additionally, we retrieved the ALMA 1.3 mm continuum map (resolution $\sim$0\rlap.{$''$}82 $\times$ 0\rlap.{$''$}58, P.A. = $-$69$\degr$.9, 1$\sigma$ $\sim$0.4 mJy beam$^{-1}$) from the ALMA science archive (project \#2017.1.01116.S; PI: Shirley, Yancy). We explored the {\it Herschel} temperature ($T_\mathrm{d}$) map \citep[resolution $\sim$12$''$;][]{molinari10b,marsh15,marsh17} of G25.4NW. The Bayesian {\it PPMAP} procedure \citep{marsh15,marsh17} was used to build the {\it Herschel} map.  
We obtained the processed mid-infrared (MIR) image at 19.7 $\mu$m (resolution $\sim$2\rlap.{$''$}5) 
of G25.4NW from the NASA/IPAC infrared science archive (Plan ID: 06\_0042; PI: Jonathan Tan), which was observed with the SOFIA Faint Object infraRed CAmera for the SOFIA Telescope \citep[FORCAST;][]{herter12} facility. 
High resolution dust continuum map at 350 $\mu$m \citep[resolution $\sim$8\rlap.{$''$}5;][]{manuel15} observed using the Second-generation Submillimeter High Angular Resolution Camera (SHARC-II) facility. The SHARC-II continuum map was smoothed by a Gaussian function with a width of 3 pixels.  

The $^{13}$CO(J =1$-$0) line data were collected from the FOREST Unbiased Galactic plane Imaging survey with the Nobeyama 45-m telescope \citep[FUGIN;][]{umemoto17} survey. The $^{13}$CO data cube was smoothed with a Gaussian function, yielding a spatial resolution of 35$''$.
\begin{table*}
\scriptsize
\setlength{\tabcolsep}{0.1in}
\centering
\caption{List of different surveys studied in this paper.}
\label{tab1}
\begin{tabular}{lcccr}
\hline 
  Survey  &  Wavelength/Frequency/line(s)       &  Resolution ($\arcsec$)        &  Reference \\   
\hline
\hline 
 Coordinated Radio and Infrared Survey for High-Mass Star Formation (CORNISH)   & 5 GHz                       & $\sim$1.5          & \citet{hoare12}\\
The HI/OH/Recombination line survey of the inner Milky Way (THOR)                             & 1--2 GHz                       & $\sim$25          & \citet{beuther16}\\
FUGIN survey &  $^{12}$CO, $^{13}$CO, C$^{18}$O (J = 1--0) & $\sim$20        &\citet{umemoto17}\\
APEX Telescope Large Area Survey of the Galaxy (ATLASGAL)                 &870 $\mu$m                     & $\sim$19.2        &\citet{schuller09}\\
{\it Herschel} Infrared Galactic Plane Survey (Hi-GAL)                              &70--500 $\mu$m                     & $\sim$5.8--37         &\citet{molinari10}\\
{\it Spitzer} Galactic Legacy Infrared Mid-Plane Survey Extraordinaire (GLIMPSE)       &3.6--8.0  $\mu$m                   & $\sim$2           &\citet{benjamin03}\\
UKIDSS near-infrared Galactic Plane Survey (GPS)       &1.25--2.2  $\mu$m                   & $\sim$0.8           &\citet{lawrence07}\\
UKIRT Wide-field Infrared Survey for H2 (UWISH2)       &2.12  $\mu$m                   & $\sim$0.8           &\citet{froebrich11}\\
UKIRT Wide-field Infrared Survey for Fe+ (UWIFE)       &1.64  $\mu$m                   & $\sim$0.8           &\citet{lee14}\\
\hline          
\end{tabular}
\end{table*}
\section{Results}
\label{sec:data}
\subsection{Physical environment of G25.4NW}
\label{sec:morph}
In this section, we explore the distribution of the dust, molecular, and ionized emission toward areas of several parsces around G25.4NW.   
\subsubsection{{\it Spitzer} and THOR maps of G25.4NW}
\label{s1sec:morph}
Figure~\ref{fig1}a displays a three-color composite map made using the {\it Spitzer} 8.0 $\mu$m (in red), 4.5 $\mu$m (in green), and 3.6 $\mu$m (in blue) images. 
The radio 1690 MHz continuum contours are also overlaid on the color-composite map, indicating the location of the H\,{\sc ii} region G25.4NW. 
The radio continuum map at 1690 MHz is obtained from the THOR survey, which provides six radio continuum maps at 1060, 1310, 1440, 1690, 1820, and 1950 MHz \citep{bihr16}. The other five THOR maps also show a similar morphology of the H\,{\sc ii} region as in the map at 1690 MHz (not presented here). 
The excess 4.5 $\mu$m emission is found in the {\it Spitzer} color composite map, showing the location of the previously 
reported EGO in the direction of the continuum source J18354S. 
No radio counterpart is seen toward J18354S in the THOR continuum maps, while the continuum source J18354N is associated with 
the H\,{\sc ii} region. 
We find the H\,{\sc ii} region G25.4NW in the catalog of THOR radio continuum sources, which is 
designated as G25.397$-$0.141 with a positive spectral index \citep[$\alpha$ $\sim$0.83;][]{bihr16}. 
The radio spectral index is determined using the radio peak fluxes at six THOR bands. 
The positive $\alpha$ value suggests the thermal free-free emission from the H\,{\sc ii} 
region G25.4NW \citep[e.g.,][]{rybicki79,longair92}. 

In each THOR continuum map, we employed the {\it clumpfind} IDL program \citep{williams94} to estimate the total flux 
density ($S_\mathrm{\nu}$) of the H\,{\sc ii} region G25.4NW, 
which is used to compute the number of Lyman continuum photons $N_\mathrm{UV}$ of the H\,{\sc ii} region. 
The values of $S_\mathrm{\nu}$ at 1060, 1310, 1440, 1690, 1820, and 1950 MHz are 
estimated to be about 2.1, 2.3, 2.5, 2.7, 2.8, and 3.0 Jy, respectively.  
We adopted the following equation to estimate the value of $N_\mathrm{UV}$ \citep{matsakis76}:
\begin{equation}
\begin{split}
N_\mathrm{UV} (s^{-1}) = 7.5\, \times\, 10^{46}\, \left(\frac{S_\mathrm{\nu}}{\mathrm{Jy}}\right)\left(\frac{D}{\mathrm{kpc}}\right)^{2} 
\left(\frac{T_{e}}{10^{4}\mathrm{K}}\right)^{-0.45} \\ \times\,\left(\frac{\nu}{\mathrm{GHz}}\right)^{0.1}
\end{split}
\end{equation}
\noindent where $S_\mathrm{\nu}$ (in Jy) is defined earlier, 
$D$ is the distance in kpc, $T_{e}$ is the electron temperature, and $\nu$ is the frequency in GHz. 
In this calculation, we considered an electron temperature of 10000~K and $D$ = 5.7 kpc. 
The values of $\log{N_\mathrm{UV}}$ of the H\,{\sc ii} region at 1060, 1310, 1440, 1690, 1820, and 1950 MHz are 
determined to be about 48.7, 48.8, 48.8, 48.8, 48.9, and 48.9 s$^{-1}$, respectively. 
It implies that the derived value of $\log{N_\mathrm{UV}}$ is similar in each THOR band. 
Following the work of \citet{panagia73}, we find that the H\,{\sc ii} region G25.4NW is excited by an O6--O7 class star, which is in agreement with the earlier published results \citep{zhu11,ai13}. 
\subsubsection{{\it Herschel}, ATLASGAL, and FUGIN maps of G25.4NW}
\label{s2sec:morph}
\begin{figure*}
\includegraphics[width=10.cm]{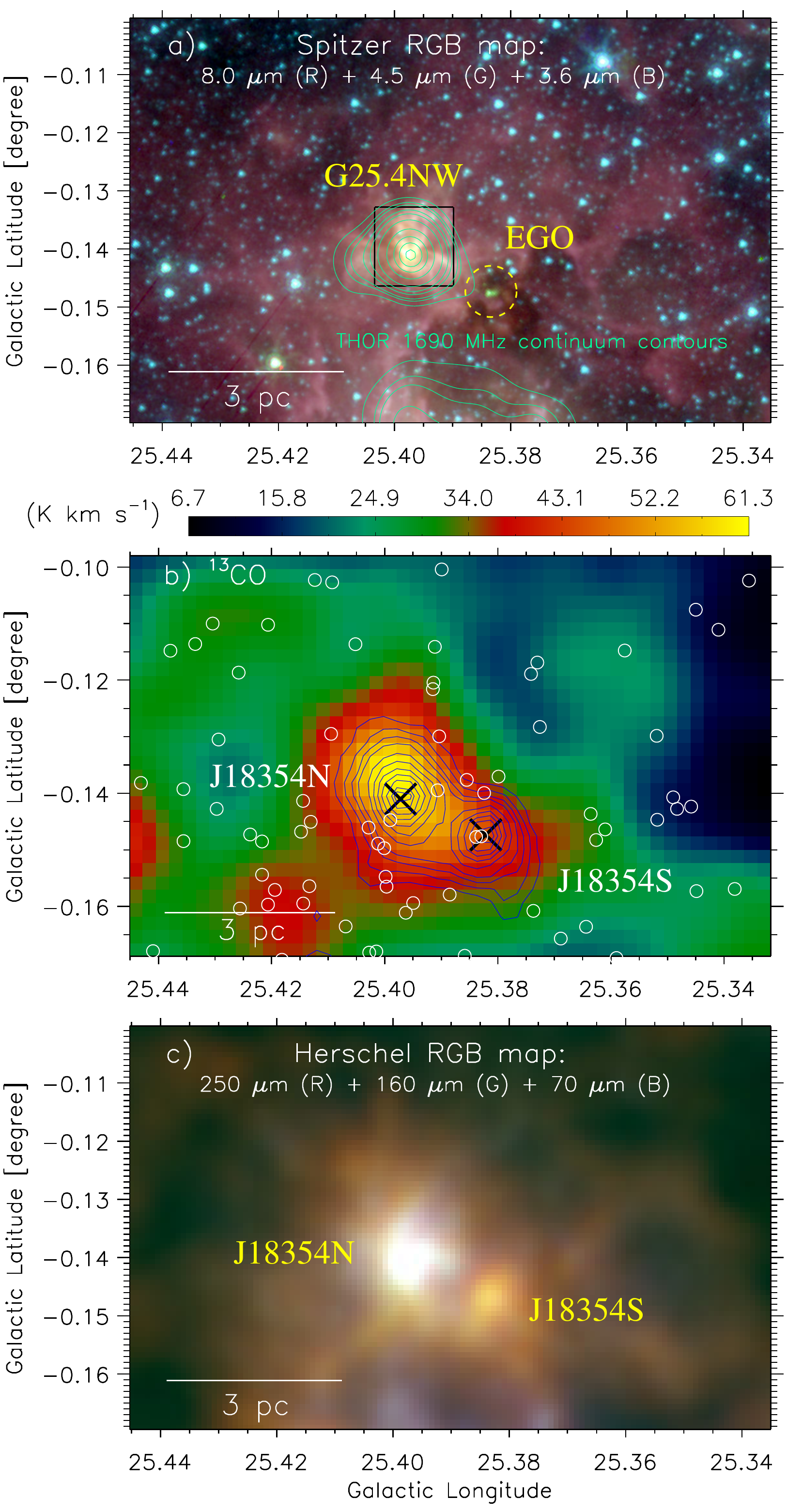}
\caption{a) Overlay of the THOR 1690 MHz radio continuum contours on the {\it Spitzer} color-composite image of 
G25.4-0.14 or G25.4NW (area $\sim$6$'$.6 $\times$ 4$'$.2; central coordinates: {\it l} = 25$\degr$.39; {\it b} = $-$0$\degr$.135) made using 8.0 $\mu$m image (red), 
4.5 $\mu$m image (green), and 3.6 $\mu$m image. The black box indicates an area shown in 
Figures~\ref{fig4}a--\ref{fig4}f.  
The radio continuum contours are plotted with the levels of (0.05, 0.08, 0.1, 0.15, 
0.2, 0.3, 0.5, 0.7, 0.85, 0.97) $\times$ 1.83 Jy beam$^{-1}$. An EGO is highlighted by a dashed circle in the figure \citep[see also Figure~4 in][]{zhu11}. b) The FUGIN intensity map of the $^{13}$CO(J =1$-$0) emission integrated 
over a velocity range of [89.5, 101.2] km s$^{-1}$. The molecular map is overlaid with 
the contours of the ATLASGAL 870 $\mu$m continuum emission and the positions of the infrared-excess sources \citep[see circles; from][]{dewangan15}. The dust continuum contours (in blue) are shown 
with the levels of (0.15, 0.2, 0.25, 0.3, 0.4, 0.5, 0.6, 0.7, 0.8, 0.9, 0.95) $\times$ 4.37 Jy beam$^{-1}$. 
Multiplication symbols show the peak positions of the ATLASGAL dust clumps. 
c) The {\it Herschel} color-composite image of G25.4NW made using 250 $\mu$m image (red), 
160 $\mu$m image (green), and 70 $\mu$m image. In each panel, the scale bar referring to 3 pc (at a distance of 5.7 kpc) is shown.} 
\label{fig1}
\end{figure*}

In Figure~\ref{fig1}b, we show the integrated $^{13}$CO intensity map at [89.5, 101.2] 
km s$^{-1}$ superimposed with the ATLASGAL 870 $\mu$m continuum contours. 
The FUGIN $^{13}$CO intensity map shows a molecular condensation hosting the dust continuum sources J18354N and J18354S. 
Note that the resolution of the FUGIN line data is insufficient to reveal inner molecular structures of these two distinct continuum sources. Using the ATLASGAL 870 $\mu$m continuum data, \citet{urquhart18} reported the total flux densities (dust temperature ($T_\mathrm{d}$)) of J18354N and J18354S to be 
$\sim$36.7~Jy~(34.4~K) and $\sim$26.6~Jy~(24.8~K), respectively \citep[see also][]{urquhart14}. However, they computed the masses of both the continuum sources at a far-side distance of 10.2 kpc. 
But, the near-side distance of the sources has been preferred in the literature \citep{zhu11,ai13,zhang19,zhang20}. Therefore, using these observed parameters and D = 5.7 kpc, we compute the masses of J18354N and J18354S to be $\sim$3150 M$_{\odot}$ and $\sim$3515 M$_{\odot}$, respectively. 
\citet{urquhart14} reported the semi-major and semi-minor axes of the ellipse bounding the clump J18354N, which are 26$''$($\sim$0.72 pc) and 21$''$($\sim$0.58 pc), respectively. 
In the case of the clump J18354S, the semi-major and semi-minor axes are 21$''$($\sim$0.58 pc) and 18$''$($\sim$0.50 pc), respectively. 

The calculation utilized the following equation \citep{hildebrand83} to compute the mass of each continuum source \citep[see also equation~1 in][]{dewangan16}:
\begin{equation}
M \, = \, \frac{D^2 \, F_\nu \, R_t}{B_\nu(T_D) \, \kappa_\nu}
\end{equation} 
\noindent where $F_\nu$ is the total integrated flux (in Jy), 
$D$ is the distance (in kpc), $R_t$ is the gas-to-dust mass ratio, 
$B_\nu$ is the Planck function for a dust temperature $T_D$, 
and $\kappa_\nu$ is the dust absorption coefficient. 
In this calculation, we considered $\kappa_\nu$ = 1.85\,cm$^2$\,g$^{-1}$ at 870 $\mu$m \citep{schuller09}, $R_t$ = 100, and $D$ = 5.7 kpc.

In Figure~\ref{fig1}b, the positions of the infrared-excess sources/YSOs \citep[from][]{dewangan15} are also marked by circles. 
Due to saturation of the {\it Spitzer} 24 $\mu$m image of G25.4NW, \citet{dewangan15} employed the {\it Spitzer} 3.6--8.0 $\mu$m and the UKIDSS GPS HK photometric data to identify these YSOs. 
About a dozen of YSOs are distributed toward the molecular condensation. 
However, no intense star formation activity or clustering of YSOs is seen toward both the continuum sources J18354N and J18354S. 
But, the MSF activity is found toward both the continuum sources.

Figure~\ref{fig1}c displays a three-color composite map made using the {\it Herschel} 250 $\mu$m (in red), 160 $\mu$m (in green), and 70 $\mu$m (in blue) images. 
The continuum sources J18354N and J18354S are clearly evident in the {\it Herschel} color composite map. 
Additionally, we also notice the existence of parsec scale filaments toward J18354N.  
Figure~\ref{fig2}a displays the {\it Herschel} temperature ($T_\mathrm{d}$) map, depicting the warm 
dust emission ($T_\mathrm{d}$ $\sim$23--39~K) toward the H\,{\sc ii} region G25.4NW. 
On the other hand, the source J18354S is seen with $T_\mathrm{d}$ of $\sim$22--23.2~K. 
\subsubsection{New candidate hub-filament system in G25.4NW}
\label{s4sec:morph}
\begin{figure*}
\includegraphics[width=10.5cm]{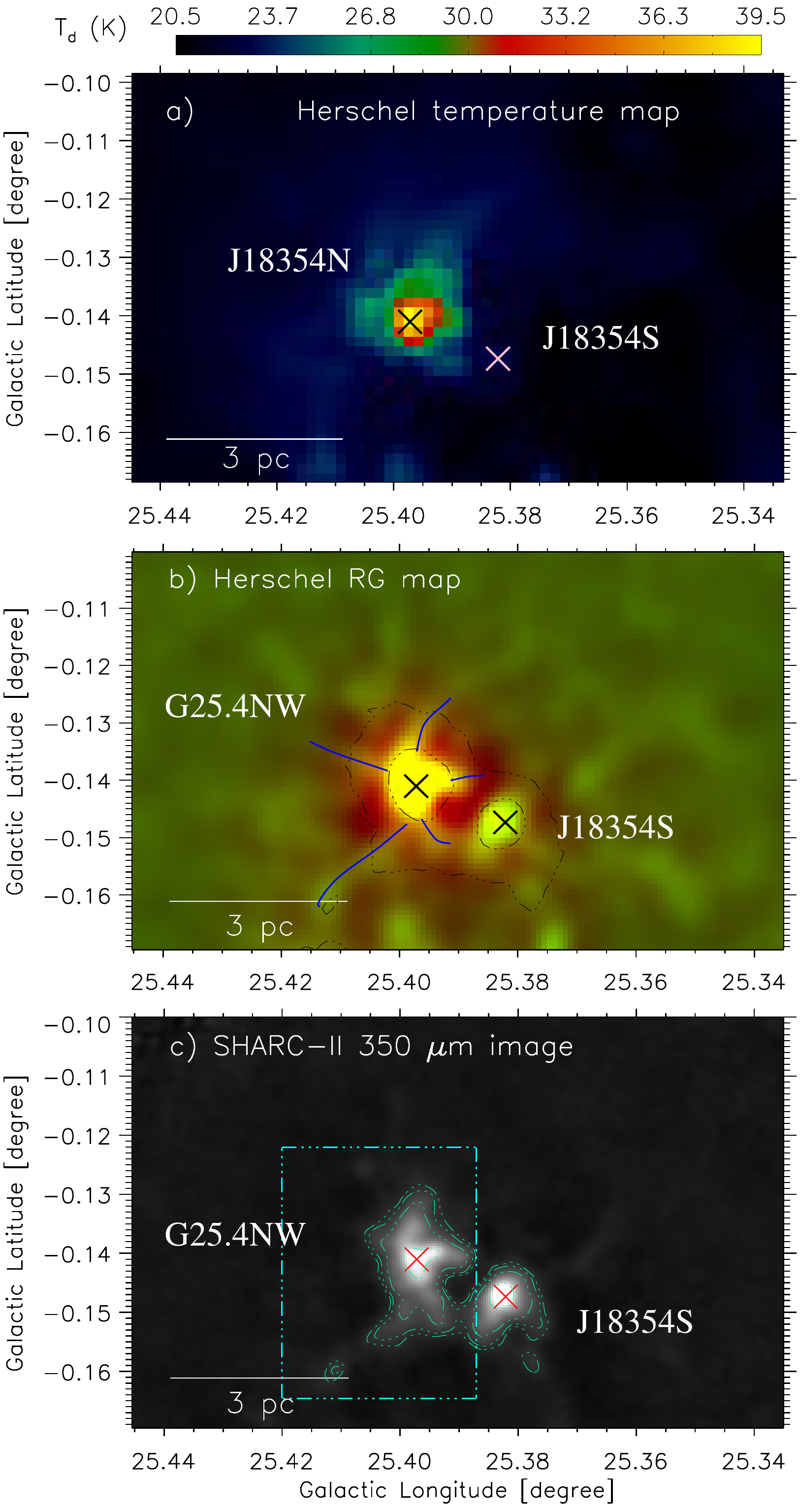}
\caption{a) {\it Herschel} temperature map of G25.4NW. 
b) Overlay of the 870 $\mu$m continuum contours on a two color-composite map ({\it Herschel} 160 $\mu$m (red) and processed {\it Herschel} 160 $\mu$m (green) images). The {\it Herschel} 160 $\mu$m image is exposed to an ``Edge-DoG" algorithm. 
The 870 $\mu$m continuum contours (in black) are also shown with the levels of (0.15, 0.5) $\times$ 4.37 Jy beam$^{-1}$. 
Solid curves (in blue) highlight filament-like features in the color-composite map. 
c) SHARC-II dust continuum map and contours at 350 $\mu$m. 
The contours (in spring green) are plotted with the levels of 1.5, 2.2, 4.1, and 13.6 Jy beam$^{-1}$.   
The continuum image is processed through a Gaussian smoothing function with a width of 3 pixels. The dotted-dashed box (in cyan) indicates an area shown in Figure~\ref{fig3x}a. 
In each panel, multiplication symbols show the peak positions of the dust clumps traced in the ATLASGAL 870 $\mu$m map 
(see also Figure~\ref{fig1}b).} 
\label{fig2}
\end{figure*}

In Figure~\ref{fig1}c, we find a hint of the presence of embedded filaments. 
In order to further explore the embedded filaments, we processed the {\it Herschel} 160 $\mu$m image through an 
edge detection algorithm \citep[i.e. Difference of Gaussian (DoG); see][]{gonzalez11,assirati14,dewangan17b,dewangan20x}. 
In Figure~\ref{fig2}b, we display a two color-composite map made using the {\it Herschel} 160 $\mu$m image 
and the DoG processed 160 $\mu$m image. The color-composite map is also overlaid with the ATLASGAL 870 $\mu$m continuum contours, and 
enables us to visually examine embedded filaments in the direction of G25.4NW or J18354N, which are indicated by solid curves.   
Based on the observed configuration of the filaments, we find the existence 
of a ``hub-filament" system in G25.4NW \citep[e.g.,][]{myers09,schneider12,baug15,dewangan15,dewangan18,dewangan20x}. 

To further explore a hub-filament system in G25.4NW, the SHARC-II 350 $\mu$m continuum map and contours are presented in Figure~\ref{fig2}c. 
We also mark the peak positions of the dust clumps traced in the ATLASGAL 870 $\mu$m map (see multiplication symbols in Figure~\ref{fig2}c). 
The SHARC-II sub-mm image has higher spatial resolution than the {\it Herschel} sub-mm images at 160--500 $\mu$m, allowing us to gain further insight into the spatial morphology of each dust clump. 
Using the SHARC-II sub-mm image, a zoomed-in view of an area around G25.4NW is presented in Figure~\ref{fig3x}a.  
The H\,{\sc ii} region G25.4NW or J18354N is exclusively traced at the central hub, 
and five parsec scale filaments seem to be converging on this hub. 
By eye examination of the SHARC-II image reveals the compact hub with low aspect ratio (length/diameter) 
and filaments with high aspect ratio. 
None of these filaments appear to be spatially connected with the continuum source J18354S (see contours in Figure~\ref{fig2}c). 
However, both the continuum sources are well located within the outer contours of 
the 350 and 870 $\mu$m continuum emission and the molecular condensation (extent $\sim$3.7 pc; see Figures~\ref{fig1} and~\ref{fig2}). Note that the hub-filament system is not resolved in the ATLASGAL 870 $\mu$m continuum map and the FUGIN $^{13}$CO data (see Figure~\ref{fig1}b). 
Hence, new molecular line observations with a higher resolution (below 10$''$) will be helpful to explore the gas motions toward the hub-filament system.

Overall, we investigate a new candidate hub-filament system in H\,{\sc ii} region G25.4NW, which is not yet reported in the literature \citep[e.g.,][]{zhu11,ai13,kumar20}.  
\subsubsection{Ionized clumps in the hub}
\label{zs1c:morph}
\begin{figure*}
\includegraphics[width=\textwidth]{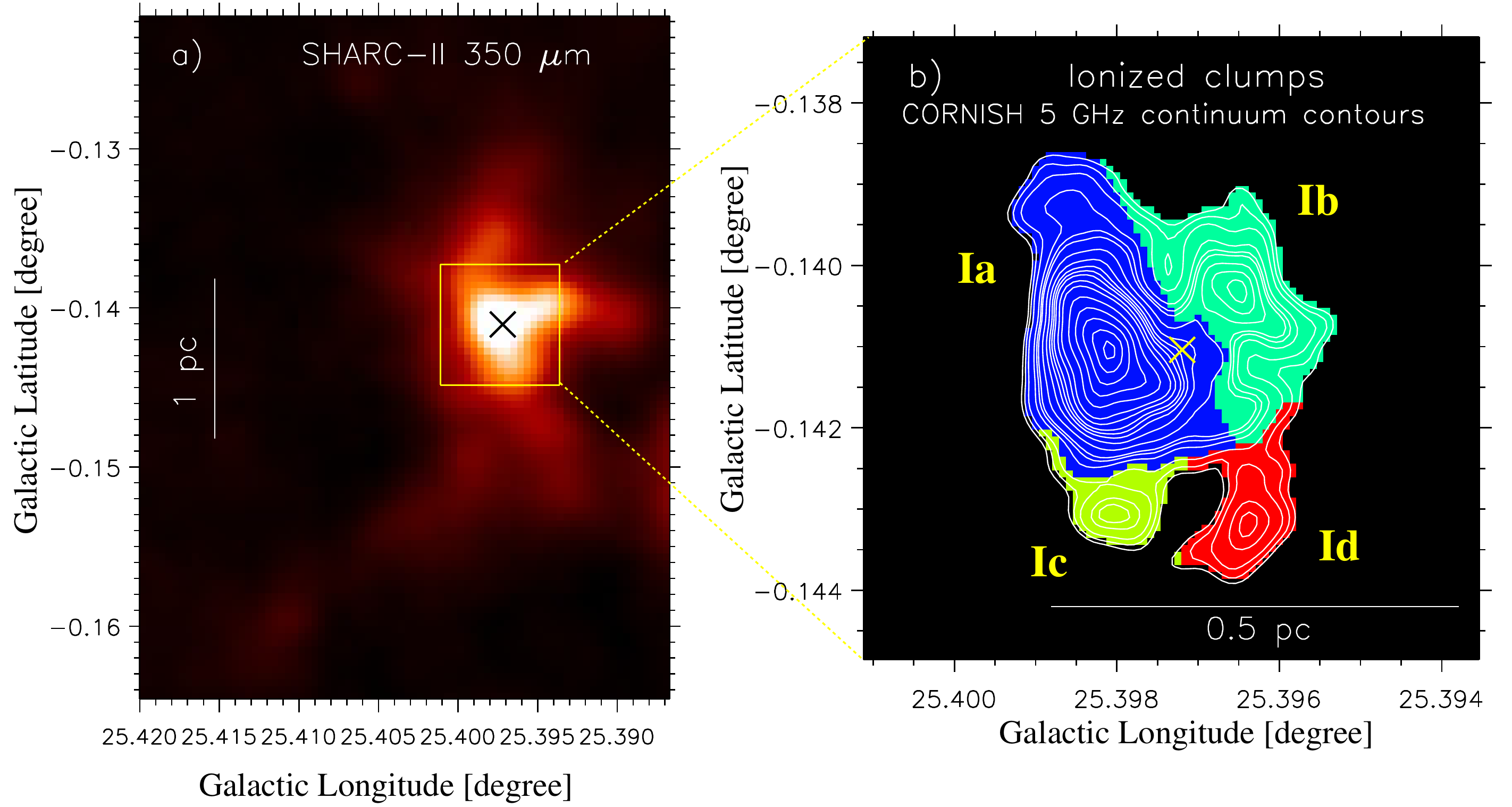}
\caption{a) A zoomed-in view of an area around G25.4NW using the SHARC-II dust continuum image at 350 $\mu$m (see a dotted-dashed box in Figure~\ref{fig2}c), 
revealing a candidate hub-filament system. b) The panel displays the CORNISH 5~GHz continuum emission of an area highlighted by a solid box in Figure~\ref{fig3x}a. 
The 5~GHz continuum contours (in white) are shown with the levels of (0.058, 0.075, 0.1, 0.15, 0.18, 0.2, 0.25, 0.3, 0.34, 0.37, 0.4, 0.45, 0.5, 0.6, 0.7, 0.8, 0.9, 0.98) $\times$ 122.1 mJy beam$^{-1}$ (where 1$\sigma$ $\sim$0.4 mJy beam$^{-1}$). 
The background map shows a clumpfind decomposition of the 5~GHz continuum emission, 
tracing the spatial extension of four ionized clumps (i.e., Ia, Ib, Ic, and Id). 
In each panel, the multiplication symbol is the same as in Figure~\ref{fig1}b.} 
\label{fig3x}
\end{figure*}

In the direction of the central hub, high resolution CORNISH radio 5 GHz continuum emission contours are presented in Figure~\ref{fig3x}b. 
At least four emission peaks are detected in the 5 GHz continuum map. With the application of the {\it clumpfind} program in the 5 GHz continuum map, we computed the total flux of the H\,{\sc ii} region G25.4NW (deconvolved effective radii ($R_\mathrm{HII}$)) to be
$\sim$2 Jy ($\sim$0.2 pc). Using the equation~1, this flux density yields the value of $\log{N_\mathrm{UV}}$ to be 48.8, which corresponds to 
a powering star of spectral type O7--O6.5. This outcome is consistent with the analysis of the THOR radio continuum data (see Section~\ref{s1sec:morph}).

The {\it clumpfind} algorithm also reveals four ionized clumps in the CORNISH map, which are designated as Ia, Ib, Ic, and Id. 
The boundary of each ionized clump is also presented in Figure~\ref{fig3x}b. 
The flux densities (deconvolved angular sizes) of the ionized clumps Ia, Ib, Ic, and Id are 
about 1408 mJy (4\rlap.{$''$}5 $\times$ 6\rlap.{$''$}5), 387 mJy (4$''$ $\times$ 6$''$), 73 mJy (3$''$ $\times$ 2\rlap.{$''$}4), and 106 mJy (2\rlap.{$''$}6 $\times$ 4\rlap.{$''$}1), respectively. 
The deconvolved effective radii ($R_\mathrm{HII}$) of the ionized clumps Ia, Ib, Ic, and Id are 
about 0.15, 0.11, 0.06, and 0.07 pc, respectively. Using the equation~1, we determine $\log{N_\mathrm{UV}}$ of Ia, Ib, Ic, and Id to be 48.6, 48.0, 47.3, and 47.5 s$^{-1}$, respectively. 
In this calculation, we utilized $T_{e}$ = 10000~K and $D$ = 5.7 kpc. Following the work of \citet{panagia73}, we find that the ionized clumps Ia, Ib, Ic, and Id are excited by O8V, B0--O9.5V, B0.5--B0V, 
and B0.5--B0V type stars, respectively. Therefore, the CORNISH continuum map favours the presence of multiple massive stars (including an O-type star) in the central hub. 
\subsection{Multi-wavelength high-resolution images of the central hub}
\label{sec:morph1}
\begin{figure*}
\includegraphics[width=\textwidth]{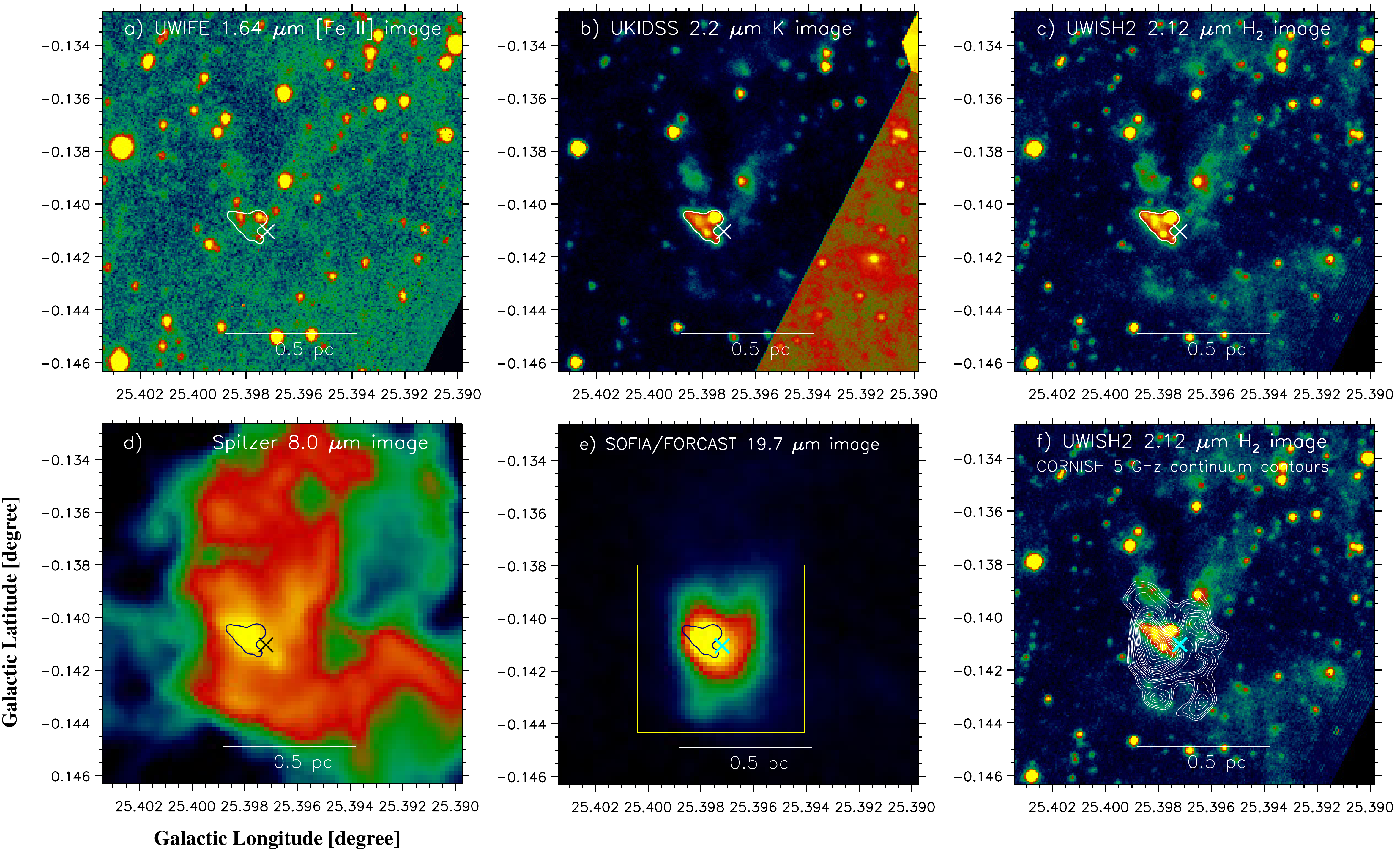}
\caption{Multi-wavelength picture of an area highlighted by a box in Figure~\ref{fig1}a. 
The continuum images are shown at different wavelengths, which are indicated in the panels. 
A compact NIR feature (extent $\sim$0.2 pc) seen in the K-band and H$_{2}$ images is indicated by a solid curve in panels ``a--e". 
In panel ``e", a solid box encompasses the area displayed in Figure~\ref{fig6}a. In panel ``f", the CORNISH 5~GHz continuum emission contours are overlaid on the H$_{2}$ image, and are the same as in Figure~\ref{fig3x}b.  
In all panels, the multiplication symbol is the same as in Figure~\ref{fig3x}a. 
In each panel, the scale bar referring to 0.5 pc (at a distance of 5.7 kpc) is displayed.}
\label{fig4}
\end{figure*}

In the direction of G25.4NW, different wavelengths are examined to explore the central hub. 
Figures~\ref{fig4}a--\ref{fig4}e show images at UWIFE 1.64 $\mu$m [Fe II] (with continuum), UKIDSS GPS 2.2 $\mu$m K-band, 
UWISH2 2.12 $\mu$m H$_{2}$ (with continuum), {\it Spitzer} 8.0 $\mu$m, SOFIA/FORCAST 19.7 $\mu$m, respectively. 
The peak position of the source J18354N traced in the ATLASGAL 870 $\mu$m map is marked by a multiplication symbol. 
In the K-band and H$_{2}$ images, we find a compact NIR feature (extent $\sim$0.2 pc), 
which is outlined by a solid curve in Figures~\ref{fig4}a--\ref{fig4}e \citep[see also Figure~3 in][]{zhu11}. 
The images at 8.0 $\mu$m and 19.7 $\mu$m are known to trace the warm dust emission. 
In Figure~\ref{fig4}f, the CORNISH radio 5 GHz continuum contours are overlaid on the H$_{2}$ image.   
The compact NIR feature is seen in the direction of the ionized clump Ia. 
Furthermore, the peak emission in the 8.0 and 19.7 $\mu$m continuum images is found toward 
the compact NIR feature, where at least three K-band objects are seen. 
Among these K-band sources, only one object appears to be detected in the 1.64 $\mu$m [Fe II] image, 
indicating the existence of embedded K-band sources. 
This argument is supported by the fact that one can easily trace embedded sources in the longer wavelength images ($>$ 2 $\mu$m), 
where lower dust extinction is expected.

\begin{figure*} 
\includegraphics[width=11.8cm]{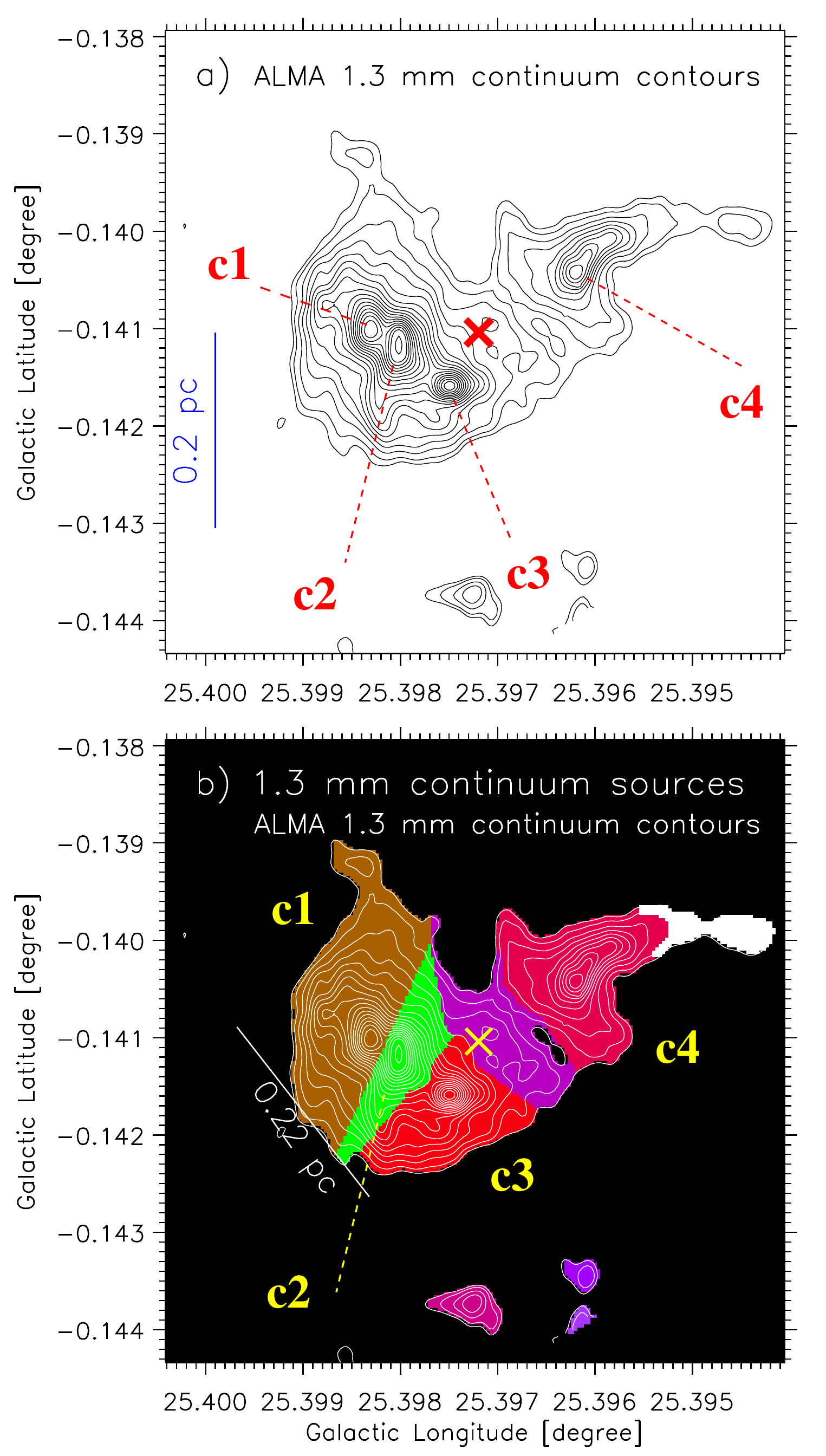}
\caption{a) A zoomed-in view of an area around G25.4NW (see a solid box in Figure~\ref{fig4}e).  
The panel displays the ALMA 1.3 mm continuum emission.
The 1.3 mm continuum contours are displayed with the levels of (0.07, 0.1, 0.15, 0.2, 0.25, 0.3, 0.35, 0.4, 0.45, 0.5, 0.55, 0.6, 0.65, 0.7, 0.75, 0.8, 0.85, 0.9, 0.98) $\times$ 25.6 mJy beam$^{-1}$ (where 1$\sigma$ $\sim$0.4 mJy beam$^{-1}$). 
b) Clumpfind decomposition of the 1.3 mm continuum emission. 
The panel shows the spatial boundaries of selected continuum sources traced in the 1.3 mm continuum image. 
The 1.3 mm continuum emission contours are also drawn in the panel (see Figure~\ref{fig6}a). 
In each panel, the multiplication symbol is the same as in Figure~\ref{fig3x}a.} 
\label{fig6}
\end{figure*}

Using the ALMA continuum emission at 1.3~mm (beam size $\sim$0\rlap.{$''$}82 $\times$ 0\rlap.{$''$}58 or 4675 AU $\times$ 3305 AU), 
a zoomed-in view of G25.4NW is presented in Figure~\ref{fig6}a (see a solid box in Figure~\ref{fig4}e). 
At least four prominent continuum peaks (i.e., c1, c2, c3, and c4) are indicated in Figure~\ref{fig6}a. 
In order to identify cores, the {\it clumpfind} program was employed in the 1.3 mm continuum map.  
At least four compact continuum sources (i.e., c1--c4) are depicted toward G25.4NW, and the boundary of each continuum source is displayed in Figure~\ref{fig6}b. 
The total flux densities (deconvolved angular sizes) of the continuum 
sources c1, c2, c3, and c4 are obtained to be about 366 mJy (2\rlap.{$''$}4 $\times$ 4\rlap.{$''$}7), 
237 mJy (1\rlap.{$''$}8 $\times$ 3\rlap.{$''$}1), 
220 mJy (2\rlap.{$''$}8 $\times$ 2\rlap.{$''$}5), and 
195 mJy (3\rlap.{$''$}1 $\times$ 2\rlap.{$''$}9), respectively. 
Using the equation~2, the masses of the continuum sources c1, c2, c3, and c4 are computed 
to be $\sim$130, $\sim$85, $\sim$80, and 70 M$_{\odot}$, respectively. 
In this estimation, we considered $\kappa_\nu$ = 0.9\,cm$^2$\,g$^{-1}$ at 1.3 mm \citep{ossenkopf94}, 
$T_D$ = 34.5 K (see Section~\ref{sec:morph}), and $D$ = 5.7 kpc. 
Based on the {\it Herschel} temperature map and the previously reported $T_\mathrm{d}$, in this paper, 
we choose the value of $T_\mathrm{d}$ $\sim$34.5~K as the representative value of J18354N. 
Using the sub-mm/mm continuum maps, the uncertainty in the mass calculation could 
be typically $\sim$20\% and at largest $\sim$50\%, which is mainly contributed by various uncertainties 
in the assumed dust temperature, opacity, and measured flux.

\begin{figure*}
\includegraphics[width=\textwidth]{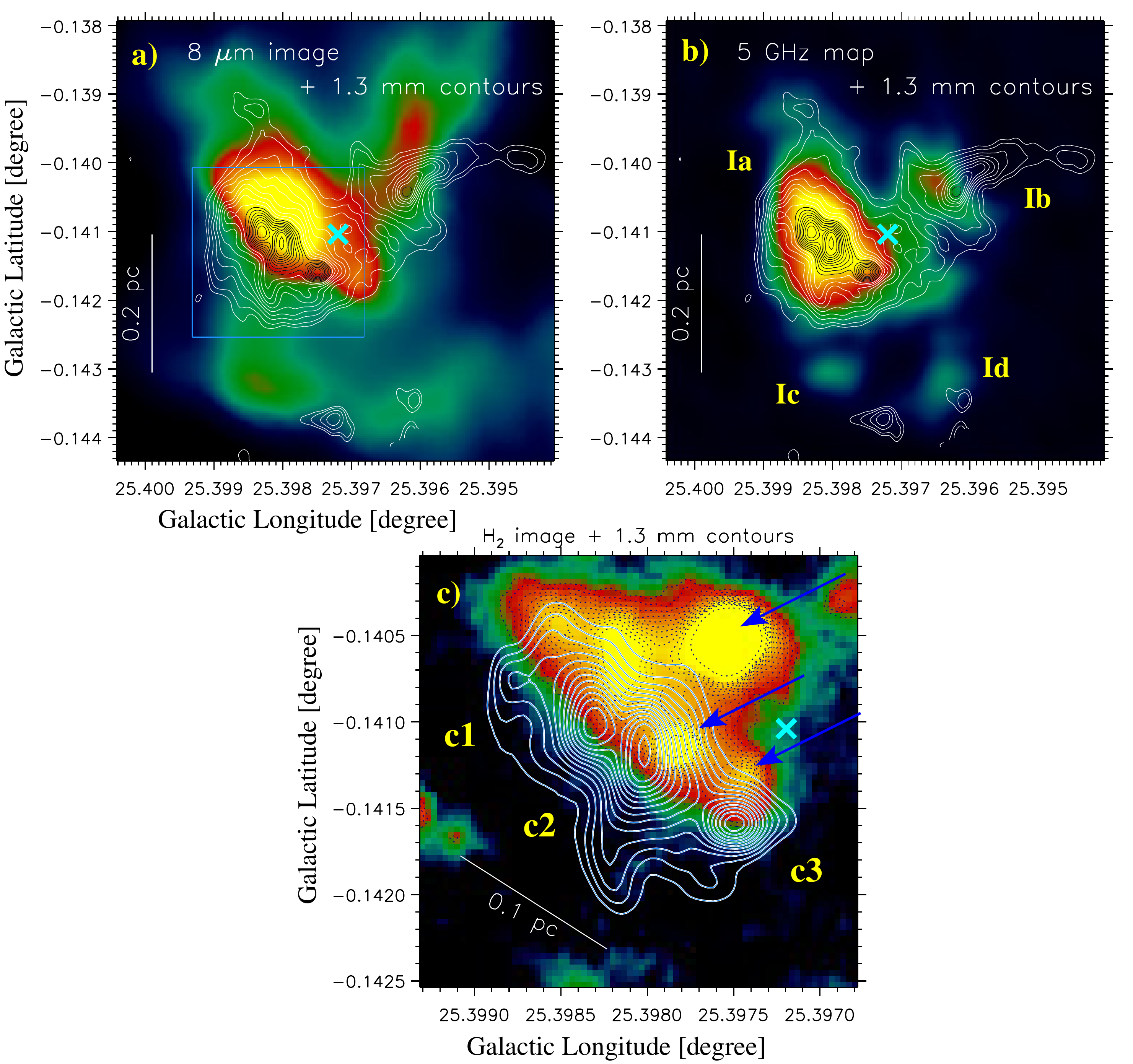}
\caption{a) Overlay of the ALMA 1.3 mm continuum emission contours on a) the {\it Spitzer} 8.0 $\mu$m image, b) the CORNISH 5 GHz continuum map. 
In panels ``a--b", the contours are the same as in Figure~\ref{fig6}a.
c) Overlay of the ALMA 1.3 mm continuum emission contours (in dodger blue) on the H$_\mathrm{2}$ image 
(see a box in Figure~\ref{fig8}a). Higher levels of the 1.3 mm continuum contours are shown in the panel (see Figure~\ref{fig6}a). 
The dotted contours (in navy) show the compact NIR feature traced in the H$_{2}$ image. Blue arrows highlight the K-band objects. In each panel, the multiplication symbol is the same as in Figure~\ref{fig3x}a.} 
\label{fig8}
\end{figure*}

A comparison of the ALMA 1.3 mm continuum emission with the continuum images at {\it Spitzer} 8.0 $\mu$m and 
CORNISH 5~GHz is presented in Figures~\ref{fig8}a and~\ref{fig8}b, respectively. 
The 1.3 mm continuum peaks (i.e., c1, c2, and c3) are found toward the ionized clump Ia, 
where the warm dust emission is traced in the MIR images. 
Additionally, the remaining peak c4 is also seen toward the CORNISH ionized clump Ib. 

In the direction of the compact NIR feature, Figure~\ref{fig8}c presents the 1.3 mm continuum emission contours overlaid on the H$_{2}$ image. 
An elongated feature (extent $\sim$0.2 pc) containing three continuum peaks (i.e., c1, c2, and c3) is evident in the 1.3 mm continuum map. The separation between peaks c1 and c2 is about 7260 AU, while the separation between 
peaks c2 and c3 is about 13390 AU. We do not find the positions of any point-like sources exactly coincident with the continuum peaks (see arrows in Figure~\ref{fig8}c). 
However, the central continuum peak, c2 is seen very close to one K-band source that is not detected in the [Fe II] image.
\section{Discussion}
\label{sec:disc}
The present paper employs several sub-mm maps of the H\,{\sc ii} region G25.4NW associated with the dust continuum source J18354N (mass $\sim$3150 M$_{\odot}$; see Section~\ref{sec:morph}). 
For the first time, we report a new picture in G25.4NW, which is a hub-filament configuration (see Section~\ref{s4sec:morph}). 
Morphologically this system consists of five parsec scale filaments, which are directed to the central hub (see Figure~\ref{fig3x}a). 
High resolution continuum maps observed in the NIR, mm, and radio wavelengths (resolution $\sim$0$''$.6--1$''$.5) have enabled us to uncover hidden substructures in the central hub. 
The presence of at least four ionized clumps (Ia--Id) excited by massive OB stars is investigated 
toward the central hub (see Figure~\ref{fig3x}b). 
The ALMA 1.3 mm continuum map detects three continuum sources c1--c3 (mass $\sim$80--130 M$_{\odot}$) and one source c4 
(mass $\sim$70 M$_{\odot}$) toward the ionized clumps Ia and Ib, respectively. 
The continuum sources c1--c3 are well distributed within an elongated feature (extent $\sim$0.2 pc) depicted in 
the ALMA 1.3 mm continuum map. 
This particular feature spatially coincides with the ionized clump Ia powered by a single O8V-type star, 
and is seen close to the compact NIR feature (extent $\sim$0.2 pc) containing embedded objects (see Figure~\ref{fig8}). 
Sub-arcsecond radio continuum observations will be helpful for further resolving the ionized clump Ia. 

It appears that there are no massive prestellar cores in J18354N. 
These observed substructures are exclusively found toward the central hub or at the center of the massive 
clump J18354N, and seem to be most closely associated with MSF. 
It may also be inferred with the detection of the warm dust emission (T$_{d}$ $\sim$23--39~K) toward 
the source J18354N, suggesting that the clump J18354N is heated 
internally by stellar feedback from massive stars (i.e., ionized emission, stellar wind, and radiation pressure). 
Collectively, the existence of a small cluster of massive stars is evident in 
the central hub of the hub-filament system in G25.4NW. 

Recently, \citet{zhang20} analyzed the NH$_{3}$ line data (beam size $\sim$4\rlap.{$''$}1 $\times$ 3\rlap.{$''$}4) toward the continuum source J18354S or G25.38 (mass $\sim$3515 M$_{\odot}$). 
However, these authors do not include an area around the source J18354N. 
In order to study the NH$_{3}$ emission toward the source J18354N, we examined this publicly available NH$_{3}$ (1,1) data cube, which covers an area containing J18354N and J18354S (not shown here). However, no NH$_{3}$ (1,1) emission is detected toward J18354N. 
This particular exercise and the observed 350 $\mu$m dust continuum emission show that the candidate hub-filament system J18354N is not a piece of the continuum source J18354S. 
As mentioned earlier, both the continuum sources are located within common ATLASGAL 870 $\mu$m contour (see Figure~\ref{fig2}b). 
It is likely that the stellar feedback from massive stars in G25.4NW may influence the environment of J18354S/G25.38. 
In this context, the knowledge of various pressure components (e.g., pressure of an H\,{\sc ii} region $(P_{\rm HII})$, 
radiation pressure (P$_{\rm rad}$), and stellar wind ram pressure (P$_{\rm wind}$)) driven by OB-type stars \citep[see][for more details]{dewangan15,dewangan17a} will be helpful to study the feedback of massive stars. However, such study is beyond the scope of this paper.
 
To explain the formation of massive stars, two groups of theoretical scenarios have been reported in the literature, which are ``core-fed" 
and ``clump-fed". In the ``core-fed" group, turbulent core accretion model explains the birth of a massive star (or a binary or a small number of multiples) via 
the collapse of a massive, isolated, and gravitationally bound prestellar core \citep{mckee03}. 
The competitive accretion and global hierarchical collapse (GHC) scenarios are part of the ``clump-fed" group \citep[see a review article by][for more details]{rosen20}.
The competitive accretion model predicts a mass accumulation through global gravitational forces in the central region of 
the clumps enclosed by smaller scale multiple cores \citep{bonnell02,bonnell04,bonnell06}. According to this model, the location of more massive stars is expected at the centre of a protostellar cluster. 
In the GHC model, one can expect gravitationally driven fragmentation in star-forming molecular clouds, 
and one can also expect large scale accretion flows to directly feed the sites of massive stars \citep[see][]{Vazquez-Semadeni+2009,Vazquez-Semadeni+2017,Vazquez-Semadeni+2019,Smith+2009}. In other words, the model predicts the fast growth of cores through accretion streams linked with the GHC of clouds \citep[e.g.,][]{Smith+2009}. 

As highlighted in Section~\ref{sec:intro}, the presence of a hub-filament configuration hints 
the onset of the global non-isotropic collapse scenario \citep[GNIC;][and see also Section~\ref{sec:intro}]{Motte+2018,Tige+2017}, which includes the basic ingredients of the GHC and clump-fed accretion models. 
Ultimately the global non-isotropic collapse scenario suggests that an infrared-bright massive protostar forms via the inflow from larger scales, and produces an H\,{\sc ii} region (in a time of few $10^{5}$--$10^{6}$~yr) through its the stellar UV radiation.
Some known sites hosting a hub-filament system are Ophiuchus \citep{myers09}, Taurus \citep{myers09}, Rosette \citep{schneider12}, IRDC G14.225$-$0.506 \citep{busquet13}, SDC335.579$-$0.292 \citep{peretto13}, Sh 2-138 \citep{baug15}, W42 \citep{dewangan15}, IRAS 05480+2545 \citep{dewangan17b}, Sh 2-53 \citep{baug18}, and Monoceros~R2 \citep{morales19}, and G18.88$-$0.49 \citep{dewangan20x}. 
More details of the global non-isotropic collapse scenario can be found in \citet{Tige+2017,Motte+2018} \citep[see also][]{morales19,dewangan20x}. 

As mentioned in this section, in G25.4NW, the central hub hosts infrared-bright massive stars and multiple cores, and are surrounded by filaments. 
Hence, the GNIC scenario appears to explain the observed morphology and MSF 
in the continuum source J18354N. 
The observed results may correspond to the infrared-bright protostellar core phase \citep[see step~5 in][]{Tige+2017}. 
However, the applicability of the turbulent core accretion model is unlikely in G25.4NW. 
This paper does not present the gas motion toward the candidate hub-filament system in G25.4NW. 
Hence, to give a conclusive interpretation concerning the target G25.4NW, high resolution 
molecular line observations will be helpful to trace the signatures of the gas flow along the filaments 
to the central hub and interactions among cores.
\section{Summary and Conclusions}
\label{sec:conc}
This paper focuses on the understanding of the process of mass accumulation in MSF. 
In this context, we present multi-scale and multi-wavelength data (including ALMA continuum maps) of the Galactic H\,{\sc ii} region G25.4NW, which is situated at a near-side distance of 5.7 kpc.
The main findings are:\\
$\bullet$ In G25.4NW, the {\it Herschel} sub-mm 160 and 250 $\mu$m continuum images (resolution $\sim$12--18$''$) 
and the SHARC-II 350 $\mu$m continuum map (resolution $\sim$8\rlap.{$''$}5) show a new candidate hub-filament system containing five parsec scale filaments and a central compact hub.\\
$\bullet$ The H\,{\sc ii} region G25.4NW with previously known thermal spectral index is traced toward the central hub, 
and is associated with the warm dust emission (at T$_{d}$ $\sim$23--39~K) inferred 
from the {\it Herschel} temperature map (resolution $\sim$12$''$).\\ 
$\bullet$ In the direction of the H\,{\sc ii} region G25.4NW, high resolution 5~GHz radio continuum map (resolution $\sim$1\rlap.{$''$}5) reveals four ionized clumps (i.e., Ia, Ib, Ic, and Id), which are powered by massive OB-type stars.\\ 
$\bullet$ A compact NIR feature (extent $\sim$0.2 pc) containing three K-band stars is investigated toward the ionized clump Ia (deconvolved size $\sim$0.12 $\times$ 0.18 pc), which is excited by an O8V-type star.\\
$\bullet$ In the direction of the ionized clump Ia, high resolution ALMA 1.3 mm continuum map (resolution $\sim$0\rlap.{$''$}82 $\times$ 0\rlap.{$''$}58) reveals an elongated feature (extent $\sim$0.2 pc) hosting three continuum sources (mass $\sim$80--130 M$_{\odot}$). These 1.3 mm continuum sources are found close to the compact NIR feature, where multiple massive stars are being formed.\\
 
Overall, our observational outcomes show the existence of a cluster of massive stars in the central hub. 
Hence, we suggest the onset of the GNIC scenario in G25.4NW, 
which successfully explains the birth of massive stars. 
\section*{Acknowledgments}
We are grateful to the anonymous reviewer for the constructive comments and 
suggestions, which greatly improved the scientific contents of the paper.  
The research work at Physical Research Laboratory is funded by the Department of Space, Government of India. 
This work is based [in part] on observations made with the {\it Spitzer} Space Telescope, which is operated by the Jet Propulsion Laboratory, California Institute of Technology under a contract with NASA. 
This publication makes use of data from FUGIN, FOREST Unbiased Galactic plane Imaging survey with the Nobeyama 45-m telescope, a legacy project in the Nobeyama 45-m radio telescope. 
This paper makes use of the following ALMA archive data: ADS/JAO.ALMA\#2017.1.01116.S. ALMA is a partnership of ESO (representing its member states), NSF (USA) and NINS (Japan), together with NRC (Canada), MOST and ASIAA (Taiwan), and KASI (Republic of Korea), in cooperation with the Republic of Chile. The Joint ALMA Observatory is operated by ESO, AUI/NRAO and NAOJ.
\subsection*{Data availability}
The CORNISH 5 GHz radio continuum data underlying this article are available from the publicly accessible CORNISH image cutout server\footnote[1]{https://cornish.leeds.ac.uk/public/img\_server.php/}.
The THOR radio continuum data underlying this article are available from the publicly accessible THOR website\footnote[2]{https://www2.mpia-hd.mpg.de/thor/Overview.html/}. 
The ALMA continuum data underlying this article are available from the publicly accessible JVO ALMA FITS archive\footnote[3]{http://jvo.nao.ac.jp/portal/alma/archive.do/}.
The FUGIN molecular line data underlying this article are available from the publicly accessible website\footnote[4]{https://nro-fugin.github.io/release/}.
The ATLASGAL 870 $\mu$m continuum data underlying this article are available from the publicly accessible ATLASGAL database server\footnote[5]{https://www3.mpifr-bonn.mpg.de/div/atlasgal/}.
The {\it Herschel}, {\it Spitzer}, SOFIA, SHARK-II continuum data underlying this article are available from the publicly accessible NASA/IPAC infrared science archive\footnote[6]{https://irsa.ipac.caltech.edu/frontpage/}.
The {\it Herschel} temperature map underlying this article is available from the publicly accessible website\footnote[7]{http://www.astro.cardiff.ac.uk/research/ViaLactea/}.
The UKIDSS GPS data underlying this article are available from the publicly accessible website\footnote[8]{http://wsa.roe.ac.uk/}.
The H$_{2}$ data underlying this article are available from the publicly accessible website\footnote[9]{http://astro.kent.ac.uk/uwish2/}.
The Fe [II] data underlying this article are available from the publicly accessible website\footnote[10]{http://gems0.kasi.re.kr/uwife/}.
%
%


\begin{thebibliography}{99}
%
\bibitem[Ai et al.(2013)]{ai13}	
Ai M., Zhu M., Xiao Li, Su H.-Q., 2013, RAA, 13, 935


\bibitem[Assirati et al.(2014)]{assirati14}
Assirati L., Silva N.~R., Berton L., Lopes A.~A., Bruno O.~M., 2014, Journal of Physics: Conference Series, 490(1), 2014

\bibitem[Baug et al.(2015)]{baug15}
Baug T., Ojha D.~K., Dewangan L.~K., Ninan J.~P., Bhatt B.~C., Ghosh S.~K., Mallick K.~K., 2015, MNRAS, 454, 4335

\bibitem[Baug et al.(2018)]{baug18}
Baug T. et al., 2018, ApJ, 852, 119

\bibitem[Benjamin et al.(2003)]{benjamin03}
Benjamin R.~A. et al., 2003, PASP, 115, 953

\bibitem[Beuther et al.(2016)]{beuther16}
Beuther H. et al., 2016, A\&A, 595, 32 

\bibitem[Bihr et al.(2016)]{bihr16}
Bihr S. et al., 2016, A\&A, 588, 97 

\bibitem[Bonnell \& Bate(2006)]{bonnell06} 
Bonnell I.~A., Bate M.~R., 2006, MNRAS, 370, 488

\bibitem[Bonnell et al.(2004)]{bonnell04} 
Bonnell I.~A., Vine S.~G., Bate M.~R., 2004, MNRAS, 349, 735

\bibitem[Bonnell et al.(2002)]{bonnell02} 
Bonnell I.~A., Bate M.~R., Clarke C.~J., Pringle J.~E., 2002, MNRAS, 323, 785

\bibitem[Busquet et al.(2013)]{busquet13}
Busquet G. et al., 2013, ApJ, 764, 26

\bibitem[Carolan et al.(2009)]{carolan09}
Carolan P.~B. et al., 2009, MNRAS, 400, 78

\bibitem[Dewangan et al.(2015)]{dewangan15}
Dewangan L.~K., Luna A., Ojha D.~K., Anandarao B.~G., Mallick K.~K., Mayya Y.~D., 2015, ApJ, 811, 79

\bibitem[Dewangan et al.(2016)]{dewangan16}
Dewangan L.~K., Ojha D.~K., Luna A., Anandarao B.~G., Ninan J.~P., Mallick K.~K., Mayya Y.~D., 2016, ApJ, 819, 66

\bibitem[Dewangan et al.(2017a)]{dewangan17a}
Dewangan L.~K., Ojha D.~K., Zinchenko I., Janardhan P., Luna, A., 2017a, ApJ, 834, 22

\bibitem[Dewangan et al.(2017b)]{dewangan17b} 
Dewangan L.~K., Ojha D.~K., Baug T., 2017b, ApJ, 844, 15

\bibitem[Dewangan et al.(2018)]{dewangan18}
Dewangan L.~K., Baug T., Ojha D.~K., Ghosh S.~K., 2018, ApJ, 869, 30

\bibitem[Dewangan et al.(2020)]{dewangan20x}	
Dewangan L. K., Ojha D. K., Sharma Saurabh, del Palacio S., Bhadari N.~K., Das A., 2020, ApJ, 903, 13

\bibitem[Froebrich et al.(2011)]{froebrich11}
Froebrich D. et al., 2011, MNRAS, 413, 480

\bibitem[Gonzalez \& Woods(2011)]{gonzalez11}
Gonzalez R, Woods R., 2011, {\it Digital Image Processing} (Pearson Education) ISBN 9780133002324

\bibitem[Herter et al.(2012)]{herter12}
Herter T.~L. et al., 2012, ApJ, 749L, 18

\bibitem[Hildebrand(1983)]{hildebrand83} 
Hildebrand R.~H., 1983, Quarterly Journal of the RAS, 24, 267

\bibitem[Hirota et al.(2018)]{hirota18}
Hirota T., 2018, Publication of Korean Astronomical Society, 33, 21.

\bibitem[Hoare et al.(2012)]{hoare12}
Hoare M.~G. et al., 2012, PASP, 124, 939 

\bibitem[Kumar et al.(2020)]{kumar20} 
Kumar M.~S.~N., Palmeirim P., Arzoumanian D., Inutsuka S.~I., 2020, A\&A, 642, 87

\bibitem[Lawrence et al.(2007)]{lawrence07}
Lawrence A. et al., 2007, MNRAS, 379, 1599

\bibitem[Lee et al.(2014)]{lee14}
Lee J.~J. et al., 2014, MNRAS, 443, 2650

\bibitem[Lester et al.(1985)]{lester85}
Lester D.~F., Dinerstein H.~L., Werner M.~W., Harvey P.~M., Evans N.~J., II, Brown R.~L., 1985, ApJ, 296, 565
 
\bibitem[Liu et al.(2011)]{liu11}  
Liu T., Wu Y., Zhang Q., Ren Z., Guan X., Zhu M., 2011, ApJ, 728, 91

\bibitem[Marsh et al.(2015)]{marsh15} 
Marsh K.~A., Whitworth A.~P., Lomax O., 2015, MNRAS, 454, 4282

\bibitem[Marsh et al.(2017)]{marsh17} 
Marsh K.~A. et al., 2017, MNRAS, 471, 2730

\bibitem[Matsakis et al.(1976)]{matsakis76}
Matsakis D.~N., Evans N.~J., II, Sato T., Zuckerman B., 1976, AJ, 81, 172

\bibitem[McKee \& Ostriker(2003)]{mckee03}
McKee C.~F., Tan, J.~C., 2003, ApJ, 585, 850

\bibitem[Manuel et al.(2015)]{manuel15}	
Merello M., Evans II N. J., Shirley Y. L., Rosolowsky E., Ginsburg A., Bally J., Battersby C., Dunham M.~M., 2015, ApJS, 218, 1

\bibitem[Molinari et al.(2010a)]{molinari10}
Molinari S. et al., 2010a, A\&A, 518, L100

\bibitem[Molinari et al.(2010b)]{molinari10b}
Molinari S. et al., 2010b, PASP, 122, 314

\bibitem[Motte et al.(2018)]{Motte+2018} 
Motte F., Bontemps S., Louvet F., 2018, ARA\&A, 56, 41 

\bibitem[Myers (2009)]{myers09} 
Myers P.~C., 2009, ApJ, 700, 1609

\bibitem[Longair(1992)]{longair92} 
Longair M.~S., 1992, High energy astrophysics. Vol.1:
Particles, photons and their detection, 436

\bibitem[Ossenkopf \& Henning(1994)]{ossenkopf94}
Ossenkopf V., Henning T., 1994, A\&A, 291, 943


\bibitem[Panagia(1973)]{panagia73} 
Panagia N., 1973, AJ, 78, 929 

\bibitem[Peretto et al.(2013)]{peretto13}	
Peretto N. et al., 2013, A\&A, 555, 112

\bibitem[Rosen et al.(2020)]{rosen20}
Rosen A.~L., Offner S.~S.~R., Sadavoy S.~I., Bhandare A., V\'azquez-Semadeni E., Ginsburg A., 2020, SSRv, 216, 62

\bibitem[Rybicki \& Lightman(1979)]{rybicki79} 
Rybicki G.~B., Lightman A.~P., 1979, Radiative processes in astrophysics

\bibitem[Schneider et al.(2012)]{schneider12}
Schneider N. et al., 2012, A\&A, 540, L11

\bibitem[Schuller et al.(2009)]{schuller09}
Schuller F. et al., 2009, A\&A, 504, 415

\bibitem[Smith et al.(2009)]{Smith+2009} 
Smith R.~J., Longmore S., Bonnell I., 2009, MNRAS, 400, 1775


\bibitem[Tig{\'e} et al.(2017)]{Tige+2017} 
Tig{\'e} J. et al.\ 2017, A\&A, 602, A77 

\bibitem[Trevi{\~n}o-Morales et al.(2019)]{morales19}
Trevi{\~n}o-Morales S.~P. et al., 2019, A\&A, 629, A81

\bibitem[Tan et al.(2014)]{tan14} 	
Tan J.~C., Beltr\'an M.~T., Caselli P., Fontani F., Fuente A., Krumholz M.~R., McKee C.~F., Stolte A., 2014, in Protostars and Planets VI, ed. H. Beuther et al. (Tucson, AZ: Univ. Arizona Press), 149

\bibitem[Umemoto et al.(2017)]{umemoto17}
Umemoto T. et al., 2017 PASJ, 69, 78

\bibitem[Urquhart et al.(2014)]{urquhart14} 
Urquhart J.~S. et al., 2014, A\&A, 568, 41

\bibitem[Urquhart et al.(2018)]{urquhart18} 
Urquhart J.~S. et al., 2018, MNRAS, 437, 1059

\bibitem[V\'azquez-Semadeni et al.(2009)]{Vazquez-Semadeni+2009} 
V\'azquez-Semadeni E., G\'omez G.~C., Jappsen A.~K., Ballesteros-Paredes J., Klessen R.~S., 2009, ApJ, 707, 1023

\bibitem[V\'azquez-Semadeni et al.(2017)]{Vazquez-Semadeni+2017} 
V\'azquez-Semadeni E., Gonz\'alez-Samaniego A., Col\'in P., 2017, MNRAS, 467, 1313

\bibitem[V\'azquez-Semadeni et al.(2019)]{Vazquez-Semadeni+2019}
V{\'a}zquez-Semadeni E., Palau A., Ballesteros-Paredes J., G{\'o}mez G.~C., Zamora-Avil\'es M., 2019, MNRAS, 490, 3061

\bibitem[Williams et al.(1994)]{williams94} 
Williams J.~P., de Geus E. J., Blitz L., 1994, ApJ, 428, 693

\bibitem[Wu et al.(2005)]{wu05} 
Wu Y., Zhu M., Wei Y., Xu D., Zhang Q., Fiege J.~D., 2005, ApJ, 628L, 57

\bibitem[Zhang et al.(2020)]{zhang20} 
Zhang C.-P., Li G.-X., Pillai T., Csengeri T., Wyrowski F., Menten K.~M., Pestalozzi M.~R., 2020, A\&A, 638, 105

\bibitem[Zhang et al.(2019)]{zhang19} 
Zhang C.-P. et al., 2019, A\&A, 627, 85

\bibitem[Zhu et al.(2011)]{zhu11}	
Zhu M., Davis C.~J., Wu Y., Whitney B.~A., Robitaille T., Peng R., 2020, ApJ, 739, 53

\bibitem[Zinnecker \& Yorke(2007)]{zinnecker07} 
Zinnecker H., Yorke H.~W., 2007, ARA\&A, 45, 481 
%
\end{thebibliography}
\end{document}